\documentclass[twocolumn,prl,aps]{revtex4}

\newcommand {\be}{\begin{equation}}
\newcommand {\ee}{\end{equation}}
\newcommand {\bea}{\begin{eqnarray}}
\newcommand {\eea}{\end{eqnarray}}
\usepackage{epsfig}

\begin{document}

\title {Magnetotransport Mechanisms in Strongly Underdoped $YBa_2Cu_3O_x$
Single Crystals}
\author{  E. Cimpoiasu, G. A. Levin, and C. C. Almasan}
\address{Department of Physics, Kent State University, Kent  OH 44242}
\author{A. P. Paulikas and B. W. Veal}
\address{Materials Science Division, Argonne National Laboratory, Argonne,
IL  60439\vspace{0.5cm}
{\rm \begin{quote}
We report magnetoresistivity measurements on
strongly underdoped $YBa_2Cu_3O_{x}$ ($x=6.25$ and
$6.36$) single crystals in applied magnetic fields $H \parallel\;$c-axis.
We identify two
different contributions to both in-plane
$\Delta\rho_{ab}/\rho_{ab}$ and out-of-plane $\Delta\rho_{c}/
\rho_{c}$ magnetoresistivities. The first contribution has the same sign as
the temperature coefficient of the resistivity
$\partial \ln\rho_{i}/\partial T$ ($i=\{c,ab\}$).  This contribution
reflects the incoherent nature of the out-of-plane transport.  The second
contribution is positive, quadratic in field, with an onset temperature that
correlates to the antiferromagnetic ordering. 
\end{quote}
\date{\today}
}
}

\maketitle

Investigation of magnetoresistance of  layered cuprates with different
levels of doping has revealed a number of effects that are  difficult to
reconcile
with the properties of conventional metals. One striking feature of the
magnetoresistivity (MR) tensor is the opposite signs of the in-plane and
out-of-plane
MR. Specifically, for a certain range of doping, and within an extensive
temperature range,
different types of cuprates exhibit the same phenomenon: the in-plane
magnetoresistivity
$\Delta\rho_{ab}/\rho_{ab}$ is positive, while the out-of-plane
magnetoresistivity
$\Delta\rho_{c}/\rho_{c}$ is negative   (see, for example,
Refs.\cite{Yan,Harris,Kimura,Heine}). These opposite signs of MR seem to
correlate
with the contrasting  temperature dependence of the respective
resistivities, namely,
metallic ($\partial\rho/\partial T>0$) in-plane resistivity $\rho_{ab}$ and
nonmetallic
($\partial\rho/\partial T<0$) out-of-plane resistivity $\rho_{c}$
\cite{Cooper}.

Another important aspect of the physics of the cuprates is the interplay
between the
charge and spin  subsystems located in the
$CuO_2$ planes.
One way to probe the charge-spin interaction across the
phase diagram is through magnetoresistance measurements.
The majority of the investigations were limited, however, to the optimally
doped or
moderately underdoped cuprates.
An investigation of the magnetoeffects of
compositions located in the vicinity of the superconducting (SC) to
antiferromagnetic
(AF) phase transition could provide important information about
the role played by the spin degrees of freedom on the
nucleation of the superconducting state.

In this paper, we address these issues through magnetoresistivity measurements
on strongly underdoped
$YBa_2Cu_3O_x$ single crystals ($x=6.25$ and $6.36$) with the magnetic
field $H$ applied parallel to the c-axis.
Our main result is  that the magnetoresistivities of these cuprates reflect
the superposition of two
contributions: (i) One contribution has the same
sign as the corresponding temperature coefficient of the
resistivity (TCR)
$\partial \ln\rho_{i}/\partial T$ ($i=\{c,ab\}$). Thus, if the in-plane
resistivity is metallic
($\partial
\ln\rho_{ab}/\partial T>0$) and the out-of-plane resistivity is nonmetallic
($\partial \ln\rho_{c}/\partial T <0$), as is the case in underdoped
crystals over an
extended temperature range, the in-plane magnetoresistivity is positive
while the
out-of-plane magnetoresistivity is negative.  The incoherence of the
out-of-plane
transport  would lead to such a correlation, of the form:
$\Delta\rho_i /\rho_i=Q\;\partial \ln\rho_{i}/ \partial T$, where either
$Q=\zeta^{orb} H^2>0$, reflecting the
conventional orbital contribution to MR in the weak field regime, or
$Q\propto \ln(H/H_0)>0$, reflecting the two dimensional 2D quantum
interference. (ii)
The other contribution to MR is positive irrespective of the sign of TCR,
correlates with the onset of
AF ordering, and has a $\gamma_{i}^{AF} H^2$ dependence. 

Magnetoresistivity measurements were carried out on two strongly underdoped
$YBa_2Cu_3O_{x}$ ($x=6.25$ and $6.36$)
 single crystals, by keeping the temperature $T$ constant while sweeping
the magnetic
field $H$ up  to $14\;T$, applied  parallel to the c-axis. Both components
of the
resistivity tensor,
$\rho_{c,ab}$, as well as both in-plane
$\Delta\rho_{ab}/\rho_{ab}$ and out-of-plane
$\Delta\rho_c/\rho_c$ magnetoresistivities were  measured by a
multiterminal method
\cite{Jiang} on the {\it same single crystal} as described in Ref.
\cite{Cimpoiasu}.
This allowed us to carry out a {\it quantitative} comparison  between
$\Delta\rho_{ab}/\rho_{ab}$  and $\Delta\rho_c/\rho_c$  as a function of
$T$  and
$H$. We note that special care was taken to maintain a constant temperature
during
the magnetic field sweep and to eliminate the Hall effect contribution to
the measured
magneto-voltages.

The puzzling coexistence of
nonmetallic $\rho_{c}(T)$ and metallic
$\rho_{ab}(T)$, characteristic to underdoped cuprates, is present in both
concentrations. The
out-of-plane resistivity is nonmetallic at all measured
$T$ for both oxygen concentrations. The in-plane resistivity
remains metallic down to
$T_{min}\approx 50\ K$ for $x=6.36$ and down to $T_{min}\approx 200\ K$ for
$x=6.25$, where
it turns insulating as well. The long-range AF ordering gives rise to an
increase in zero field
$\rho_{c}(T)$ upon cooling through $T_N$, while it has no noticeable effect
on $\rho_{ab}(T)$
\cite{Cimpoiasu2}. The
$x=6.36$ single crystal has a N\'{e}el transition temperature
$T_N\approx40\;K$ [determined from $\rho_{c}(T)$] while the $x=6.25$ single
crystal is AF at
all $T\le 300\ K$.

We had recently shown that the sign
of the out-of-plane magnetoresistivity $\Delta\rho_c/\rho_c$ of the $x=6.36$ 
single crystal measured in
a magnetic field of $14\;T$ is, for $T\geq 150\;K$, the same as the sign of the corresponding
temperature
coefficient of the resistivity $\partial \ln\rho_{c}/\partial T$
\cite{Cimpoiasu1}. This is a direct
consequence of incoherent charge transport along the c-axis.
However, $\Delta\rho_{c}/\rho_{c}$ in $14\;T$ becomes positive on approaching the
antiferromagnetic AF phase (for $T\leq
125\; K$), increasing strongly with decreasing $T$, while
$\partial \ln\rho_{c}/\partial T$ remains negative \cite{Cimpoiasu1}. A
recent report has shown
that
$\Delta\rho_{c}/\rho_{c}(T)$  of strongly underdoped
$YBa_2Cu_3O_{x}$ single crystals measured in high magnetic fields $H
\parallel\;c$-axis becomes positive well
above
$T_N$ and increases sharply with decreasing $T$ through
$T_{N}$ \cite{Lavrov}. This positive
contribution to
$\Delta\rho_{c}/\rho_{c}$ which is quadratic in $H$ was attributed to AF
correlations \cite{Lavrov}. Since
$T_N \approx 40\;K$ for the $x=6.36$ single crystal, we associate the
positive term which
increases with decreasing
$T$ for $T\leq 125\; K$, to increasing AF correlations with
decreasing
$T$.

\begin{figure}
\begin{center}
\epsfig{file=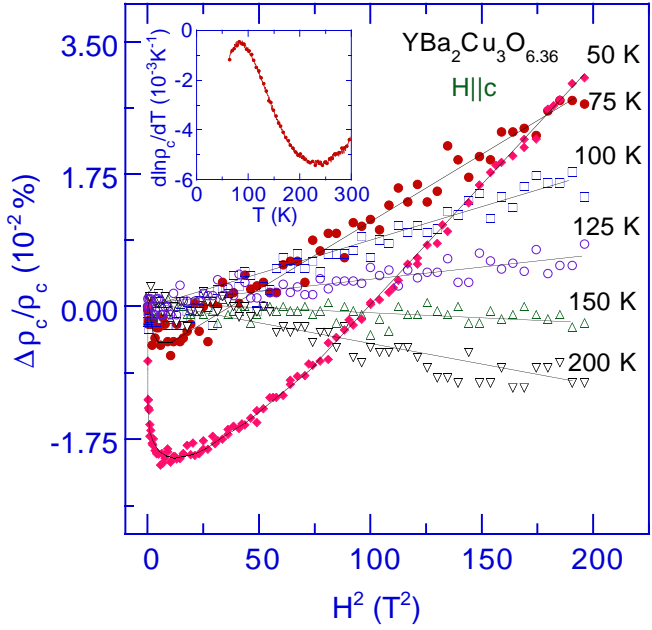,width=\columnwidth}
\end{center}
\caption{ Magnetic field $H$ dependence of the
out-of-plane magnetoresistivity $\Delta\rho_{c}/\rho_{c}$ of
$YBa_2Cu_3O_{6.36}$ single crystal.
Inset: T-dependence of the corresponding
temperature coefficient of the resistivity $\partial
\ln\rho_{c}/\partial T$ measured in $H=0\;T$.}
\end{figure}

In showing that the correlation between MR
and the corresponding TCR is a signature  of incoherent c-axis charge
transport in
these samples,
we start with the understanding that
a fundamental property of incoherent  c-axis conduction is
that the out-of-plane phase coherence length
$\ell_{\varphi ,c}$ (the average distance electrons travel between
dephasing inelastic collisions) does not
change with temperature or magnetic field \cite{Cimpoiasu1}. (The
out-of-plane incoherent transport of underdoped
$YBa_2Cu_3O_x$ was confirmed experimentaly
\cite{Kim,Basov,Marel}). Therefore, the only
length scale which determines the dissipation and can change with
temperature or  applied magnetic field is the
in-plane phase coherence length
$\ell_{\varphi ,ab }$.
Under these conditions, both conductivities depend only on the variable $
\ell_{\varphi ,ab }$ [$\sigma_{ab} (\ell_{\varphi ,ab })
$ and
$\sigma_{c} (\ell_{\varphi ,ab })$],  so that their temperature and field
dependences  come from that of $ \ell_{\varphi ,ab}$.  Hence:

\begin{equation}
\frac{\partial \rho_{c,ab}}{\partial H}=Q\frac{\partial
\rho_{c,ab}}{\partial T};\;\; Q\equiv
\frac{\partial\ell_{\varphi ,ab }/\partial H}{\partial\ell_{\varphi
,ab}/\partial T}.
\end{equation}
The immediate consequence is that the sign of each component of
magnetoresistivity  is
given by the sign of the corresponding TCR
since $\partial\ell_{\varphi ,ab }/\partial H<0$
\cite{Abrikosov,Altshuler} and $\partial\ell_{\varphi ,ab}/\partial T<0$
\cite{Abrikosov,Lee}.

The $H$-dependence of
$\Delta\rho_{c}/\rho_{c}$ is shown in Fig. 1 for $50\;K \leq T
\leq 200\;K$. Its inset shows the T-dependence of the temperature
coefficient of the resistivity $\partial
\ln\rho_{c}/\partial T$ (TCR) measured in zero field. (The fact that $\partial
\ln\rho_{c}/\partial T < 0$ over the whole $T$ range reflects the semiconducting
nature of  the c-axis conduction for all measured $T$.)  For $T \geq
150\;K$, the direct correlation
between the signs of MR (negative) and the corresponding TCR (negative, see
inset), i.e. Eq. (1), holds for all
the values of the applied magnetic fields. Hence,
$\Delta\rho_{c}/\rho_{c}$ is given by:
\be
\frac{\Delta \rho_{c}}{\rho_{c}}(H,T)=Q(H)\frac{\partial
\ln\rho_{c}}{\partial T}(T),
\end{equation}
with $Q \propto H^2>0$. According to Eq. (1), the $H$ dependence of $Q$,
hence, of  both
magnetoresistivities is given by the $H$ dependence of
$\ell_{\varphi,ab }$. Thus, this $H^2$ dependence of $\Delta
\rho_{c}/\rho_{c}$ observed at $T \ge 150\;K$, indicative
of weak field regime, is a result of the conventional orbital change of
$\ell_{\varphi ,ab }$ due to an
applied magnetic field $H\parallel c$; i.e., $Q=\zeta^{orb}H^2$.

For temperatures $100\;K \leq T < 150\;K$,
$\Delta\rho_{c}/\rho_{c}$ becomes positive, while
$\partial \ln\rho_{c}/\partial T$ is still negative (see inset to Fig. 1). The H
dependence is, however, still quadratic. Presumably, there are two
contributions to MR in this $T$ range: a {\it
negative} $\zeta^{orb}(\partial
\ln\rho_{c}/\partial T) H^2$ contribution which is a result of conventional
orbital contribution to MR,
described above, and a {\it positive}
$\gamma_{c}^{AF} H^2$ contribution attributed to spin-spin correlations.
This latter contribution dominates
$\Delta \rho_{c}/\rho_{c}$ at these temperatures
[$\gamma_{c}^{AF}>\zeta^{orb}(\partial
\ln\rho_{c}/\partial T)$] since $\Delta\rho_{c}/\rho_{c}$ is positive.

\begin{figure}
\begin{center}
\epsfig{file=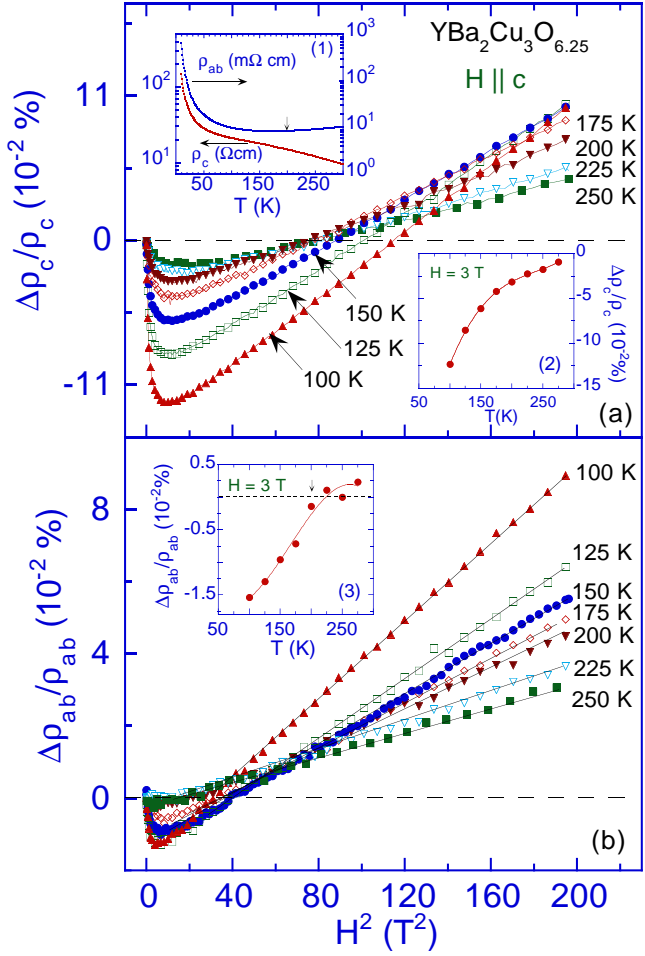,width=\columnwidth}
\end{center}
\caption{Magnetic field $H$ dependence of the (a)
out-of-plane $\Delta\rho_{c}/\rho_{c}$ and (b) in-plane
$\Delta\rho_{ab}/\rho_{ab}$ magnetoresistivities of
$YBa_2Cu_3O_{6.25}$ single crystal measured at different temperatures $T$.
Inset: (1) Temperature $T$
dependence of in-plane $\rho_{ab}$ and out-of-plane $\rho_c$ resistivities
and $T$ dependence of
(2) $\Delta\rho_{c}/\rho_{c}$ and (3) $\Delta\rho_{ab}/\rho_{ab}$ measured
at $3\;T$.}
\end{figure}

At even lower temperatures ($T=50$ and $75\;K$), a negative component in the
H-dependence of  $\Delta\rho_{c}/\rho_{c}$ is present at low fields,
superimposed on the
positive and quadratic in $H$ term. This negative sign of
$\Delta\rho_{c}/\rho_{c}$ is the same as the sign
of $\partial
\ln\rho_{c}/\partial T$. Therefore, at low $H$, $\Delta\rho_{c}/\rho_{c}$
is also given by Eq. (2), with the
$H$ dependence of the negative contribution compatible to $\ln (H/H_0)$
($H_0$ is a small characteristic
field). At low $H$, this $\ln
(H/H_0)$ contribution to $\Delta\rho_{c}/\rho_{c}$ dominates the
quadratic in $H$ contributions
(orbital and antiferromagnetic contributions)
\cite{Cimpoiasu}.  At
$H \geq 3\;T$, the $\ln
(H/H_0)$ contribution saturates while the antiferromagnetic contribution
$\gamma^{AF}H^2>0 $ takes over and
changes the sign
of $\Delta\rho_{c}/\rho_{c}$ to positive.

We had recently shown that
the $H$ dependence of
$\ell_{\varphi ,ab }$, hence, $Q$ of crystals with two-dimensional 2D phase
coherent paths is given at low enough
$T$ and $H$ by
\cite{Cimpoiasu}:

\begin{eqnarray}
 Q\propto \frac{\Delta\ell_{\varphi ,ab}}{\ell_{\varphi ,ab}}
\approx  & \left\{
\begin{array}{ll}-\eta\;\frac{\ln (H/H_0)}{\ln(H_1/H_0)}  &  H_0< H< H_1. \\
 -\eta & H>H_1.
\end{array}
          \right.
\end{eqnarray}

Here $\eta$ is a positive constant,
$H_0\sim
\phi_0/
\ell_{\varphi ,ab }^2$, and
$H_1\sim \phi_0/ l_{el}^2$, where $\phi_0$ is the magnetic flux quantum and
$l_{el}$ is the
characteristic elastic length. Therefore, the $\ln (H/H_0)$
dependence of $\Delta\rho_{c}/\rho_{c}$ observed in Fig. 1 at low
$T$ and relatively small $H$ indicates 2D quantum
interference, thus, reflects again the incoherent c-axis conduction.

Hence, for $T < 150\; K$,
there are two contributions to MR: a contribution which has the same sign
as the corresponding TCR and a
positive contribution attributed to AF correlations; i.e.,
\be
\frac{\Delta \rho_{c}}{\rho_{c}}(H,T) = Q(H)\frac{\partial
\ln\rho_{c}}{\partial T}(T) +\gamma_{c}^{AF}(T) H^2,\;\;\;
\nonumber \\
\ee
where $Q =\zeta^{orb} H^2>0$ for $T > 75\; K$ and is given
by Eq. (3)
for $T= 50 \;$ and $\; 75\;K$, and $\gamma _{c}^{AF}>0$.

The $H$ dependence of $\Delta
\rho_{ab}/\rho_{ab}$  (data not shown) is similar with the $H$ dependence of
$\Delta
\rho_{c}/\rho_{c}$ shown in Fig. 1. This reflects the fact that the $H$
dependence of both $\Delta
\rho_{c}/\rho_{c}$ and $\Delta
\rho_{ab}/\rho_{ab}$ is given by $\ell_{\varphi ,ab }(H)$. Specifically,
$\Delta\rho_{ab}/\rho_{ab}$ is positive and quadratic in $H$ over the whole
field
range for $T> 75\;K$. At $T=50$ and $75\;K$,
$\Delta\rho_{ab}/\rho_{ab}$ is {\it negative} for $H<3\;T$ and increases
quadratically for
$H>3\;T$.  The negative value of
$\Delta\rho_{ab}/\rho_{ab}$ at low fields correlates with the upturn in
$\rho_{ab}(T)$ around
$50\;K$ and, hence, with the change of sign  of $\partial
\ln\rho_{ab}/\partial T$ from positive to negative.

\begin{figure}
\begin{center}
\epsfig{file=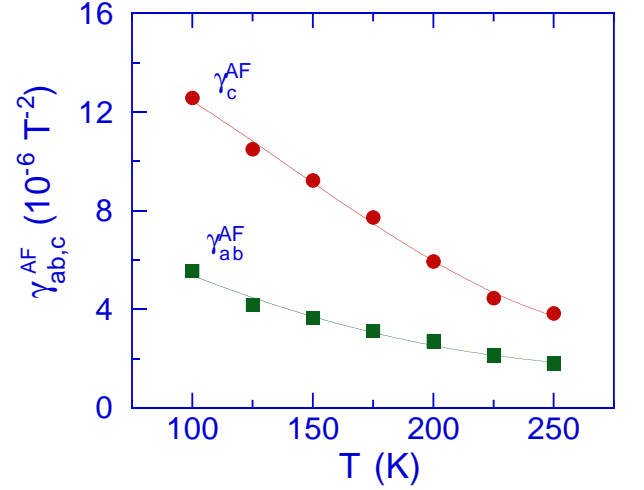,width=\columnwidth}
\end{center}
\caption{Temperature $T$ dependence of the coefficient
$\gamma_{c,ab}^{AF}$ of the quadratic magnetic
field
$H$ dependence of the magnetoresistivities of $YBa_2Cu_3O_{6.25}$ single
crystal.}
\end{figure}

To study the transport in the AF state, we also measured the
magnetoresistivity tensor of another single crystal
with a lower oxygen content ($x=6.25$) with $T_N> 300\;K$.
Figure 2(a) shows the $H$ dependence of $\Delta
\rho_c / \rho_c$ for the $x=6.25$  single crystal measured at several
temperatures.
For all temperatures $100\;K \leq T \leq 250\;K$, these curves have the
same $H$ dependence as the ones
for the
$x=6.36$ sample measured close to $T_N$
($T=50$ and $75\;K$); i.e., $\Delta
\rho_{c}/\rho_{c}$ has a nonmonotonic field dependence consistent with $\ln
H/H_0>0$ ($H_0\approx 0.3
\;T$ at $100\;K$) at low $H$ which saturates to a certain negative value
$\epsilon_c$
(for example, $\epsilon_c \approx -0.14 \%$ at $100\ K$) for
$H > 3\ T$, and a positive contribution quadratic in $H$ that takes
over at $H > 3\ T$ and
changes the sign
of $\Delta\rho_{c}/\rho_{c}$ to positive at higher fields.

The  $H$ dependence of $\Delta \rho_{ab}/\rho_{ab}$ [Fig.
2(b)] is  similar with the $H$ dependence of $\Delta \rho_c/\rho_c$ [Fig.
2(a)]. However,
the negative term in $\Delta
\rho_{ab}/\rho_{ab}$ is   about seven times smaller than the negative term
in $\Delta \rho_{c}/\rho_{c}$.

The $T$-profiles of
$\Delta \rho_c/\rho_c$ and
$\Delta \rho_{ab}/\rho_{ab}$ at $H=3\;T$, where minima in the
magnetoresistivities
occur, are plotted in inset 2 and 3, respectively, to Figs. 2. Note that
the sign of both
$\Delta \rho_c/\rho_c$ and $\Delta \rho_{ab}/\rho_{ab}$ at this low $H$ is
the same as the sign of
the corresponding TCR [see inset 1 to Fig. 2(a) which gives
$\rho_c(T)$ and $\rho_{ab}(T)$]. Indeed, for all  temperatures $100\;K \leq
T \leq 275\;K$,
$\Delta \rho_c/\rho_c$ at $H=3\;T$ is negative and increases in magnitude
with decreasing temperature. On the
other hand, $\Delta \rho_{ab}/\rho_{ab}$ at $H=3\;T$ is positive for
$T>200\;K$ and negative for $T<200\;K$.
Therefore, both magnetoresistivities of this sample are given by Eq. (4)
over the whole measured $T$ range with
$Q$ given by Eq. (3).

The coefficients $\gamma_{c}^{AF}$ and $\gamma_{ab}^{AF}$,
determined from the fit of the quadratic dependence of $\Delta \rho_{c,ab}/\rho_{c,ab}$ at high $H$ [see Eq.
(4)] for the $x=6.25$ sample, scale for $100\;K \leq T \leq 250\;K$ with
$\gamma_{c}^{AF} /\gamma_{ab}^{AF}\approx 2$. We found the same
proportionality between the two coefficients for the $x=6.36$  single crystal, but
only for the lowest measured temperature of $T = 50\;K$, presumably because the AF correlations
are strong enough at this $T$ so that the AF contribution dominates the orbital one. This scaling of
$\gamma_{c}^{AF}$ and $\gamma_{ab}^{AF}$ is a strong indication that the same mechanism is responsible for the
positive, quadratic contributions ($\gamma_{c,ab}^{AF}H^2$) to $\Delta
\rho_{c,ab}/\rho_{c,ab}$ even though, as discussed above, this effect is noticeable
weaker on the in-plane transport than on the out-of-plane transport. 

The coefficients $\gamma_{c}^{AF}$ and $\gamma_{ab}^{AF}$ for the $x=6.25$ sample are shown
in Fig. 3. Both coefficients increase with decreasing temperature for
$100\;K \leq T \leq 250\;K$.
Previous work showed that, immediately below the N\'{e}el temperature,
$\gamma_{c}^{AF}$ decreases with decreasing $T$ \cite{Lavrov}. Hence, our
data indicate
that the temperature behavior of
$\gamma_{c}^{AF}$ far into the AF regime is different from the one in the
vicinity of
$T_N$.

In summary, all the results presented indicate that both magnetoresistivities
of underdoped
$YBa_2Cu_3O_{x}$ ($x=6.36$ and $6.25$) are described by Eq. (2) above
$T_N$,  and by
Eq. (4) around and below
$T_N$.
Therefore, both in-plane and out-of-plane magnetoresistivities of strongly
underdoped  single crystals of
$YBa_2Cu_3O_{x}$ ($x=6.25$ and $6.36$) are a result of two contributions:
one that correlates in
sign and temperature dependence with
the corresponding temperature coefficient of the resistivity
${\partial ln \rho}/{\partial T}$ and has either a $\zeta^{orb}H^2$ or an
$\ln(H/H_0)$
dependence, and another one which is positive, has a
$\gamma_{c,ab}^{AF}H^2$ dependence, and
dominates at high magnetic fields ($H>3\;T$).  The first contribution is a
fingerprint of the incoherent
nature of the out-of-plane charge transport.  The second contribution reflects the
AF correlations.

{\it The authors thank  Viorel Sandu and Almut Schroeder for the critical
reading of the
manuscript. This research was supported by
the National Science Foundation under Grant No. DMR-9801990 at KSU and the
US Department of Energy under
Contract No. W-31-109-ENG-38 at ANL}.

\end{document}